\documentclass{elsart3}

 \usepackage{graphicx}

\usepackage{amssymb}

\begin{document}

\begin{frontmatter}



\title{Electrical conductance at initial stage \ in epitaxial growth of Pb \\
on modified Si(111) surface}


\author{Z. Korczak \corauthref{cor1}}
\corauth[cor1]{Corresponding author, tel.: +48 81 537 6132,
               fax: +48 81 537 6191}
\ead{korczak@tytan.umcs.lublin.pl} and
\author{T. Kwapi\'nski}
\ead{tomasz.kwapinski@umcs.lublin.pl}

\address{Institute of Physics, M. Curie-Sk\l odowska
University, \\ pl. M. Curie-Sk\l odowskiej 1, 20-031 Lublin, POLAND}

\begin{abstract}

The electrical conductance and RHEED intensities as a function of the coverage have been measured during Pb depositions at 105 K
on Si(111)-(6x6)Au with up to 4.2 ML of annealed Pb. The experiments show the strong influence of used substrates on the
behavior of the conductance during the epitaxy of Pb atoms, especially for very initial stage of growth. Oscillations of the
conductance during the layer-by-layer growth are correlated with RHEED intensity oscillations. The analysis of the conductance
behavior is made according to the theory described by Trivedi and Aschcroft (Phys.Rev.B 38,12298 (1988)).

\end{abstract}


\end{frontmatter}


\section{\label{sec1}Introduction}
The transport properties of a clean, well ordered surface with deposited unreactive metals are an important topic of surface
physics. Since the detailed surface structure may modify the transport mechanism dramatically, the atomic structure and
transport properties have to be measured simultaneously. Most studies of electronic transport in such structures have been
conducted for metals coverage higher than 1 monolayer (ML) [1-5].  Ja{\l}ochowski \emph{et al.} concentrated on quantum size
effect studies in thin epitaxial Pb and Pb-In films grown on Si(111) reconstructed surface [1,2]. Pfennigstorf \emph{et al.}
[3,4] studied electronic transport in ultrathin epitaxial Pb films on Si(111)-7x7 and on Si(111)-$\sqrt{3}$x$\sqrt{3}$ Pb
surfaces. These authors obtained useful information about conduction mechanisms for the coverage higher than 1 ML. They did not
mention the initial decreasing in the conductance for the coverage lower than 1 ML. Hasegawa and Ino monitored the conductance
dependence on the substrate-surface structures and epitaxial growth modes at initial stages of Ag and Au depositions on a
Si(111) surface at room temperatures [5]. Authors paid little attention to the conductance changes for the coverage lower than 1
ML. Recently, Pfennigstorf \emph{et al.} [6] have concentrated on the correlation of structural properties with the measured
conductance. The conductance of a Pb film during deposition at 15 K on a film of tenths ML, which had been annealed
(recrystallized), showed the strong decrease during the first half monolayer. The behavior of the conductivity as a function of
film thickness is a result of the competition between the thickness increment and the roughness variation. Also the electron
mean free path as well as the Fermi energy and electron density are changed (oscillate) during the layer-by-layer growth but for
thicker films these functions tend to constant values which correspond to the bulk material \cite{tri,wang}. Understanding of
the electron transport characteristics for low coverage is still limited.

In this paper we focus on the measurements of the electrical conductance at initial stage in epitaxial growth of Pb on
Si(111)-(6x6)Au/Pb  (i.e. Si(111)-(6x6)Au with up to 4.2 ML of annealed Pb) surface using four-point probe. By relating these
measurements to the observed simultaneously reflection high energy electron diffraction (RHEED) intensity changes we have
obtained a better understanding of the origin of the conductance behavior. The analysis of the conductance behavior is made
according to the theory described by Trivedi and Aschcroft \cite{tri}.

\section{\label{sec2}Experimental Setup and Results}
The measurements were performed in a UHV chamber with a base pressure in order of 5x$10^{-11}$ Torr. The structure of the
substrate and the deposition of Pb were monitored by the RHEED system. An n-type Si(111) wafer of around 25 $\Omega$cm
resistivity and 18x4x0.4mm$^3$ size was mounted on a pair of  Mo rods and clamped with Ta stripes. Electrical conductivity was
measured in situ by four-point probe method. An alternating current: I = 2$\mu$A, 17 Hz was sent through the outer-most Ta
clamps contacts, while the voltage ac was measured across the inner two W wires kept in elastic contact with the wafer. The
voltage electrode spacing of 1.5 mm were used in the experiments. The sample holder was mounted to SuperTran-VP continuous flow
cryostat cold finger (Janis Research Comp. Inc.). The temperature of the sample has been measured with the Au-Fe chromel
thermocouple in touch with a wafer. Before each measurement run, the surface was cleaned to obtain a clear Si(111)-7x7 RHEED
pattern, by few flash heating for 5 seconds  with a direct current of 13.5A through the sample. In order to prepare the
Si(111)-(6x6)Au surface structure, 1.3 ML of Au were deposited on Si(111)-7x7 superstructure. Annealing for 1 min. at about 950
K and slow cooling to room temperature (10K/min.) resulted in the appearance of a sharp (6x6)Au superstructure RHEED pattern.
Si(111)-(6x6)Au with up to 4.2 ML of Pb deposited at 105K have been heated up to 300K in order to increase crystalline order.
After recrystallization the surface has been cooled to about 100K for further deposition of Pb. The conductance has been
measured simultaneously with deposition. The amount of deposited material in units of monolayer (ML = 7.8x$10^{14}$
atoms/cm$^2$) was monitored with a quartz crystal oscillator. RHEED specular beam intensity oscillation during
monolayer-by-monolayer growth has been used for calibration. Pb was evaporated from Ta crucible at rate 0.02 ML/min.

\begin{figure}[h]
\begin{center}
 \resizebox{1.0\columnwidth}{!}{
  \includegraphics{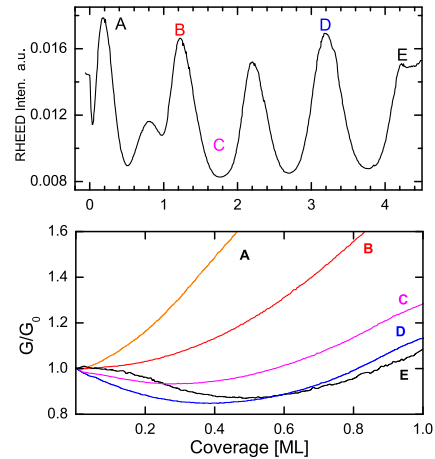}}
\end{center}
 \caption{\label{Fig1} (color online) RHEED specular intensity (upper panel) during growth of
 Pb on Si(111)-(6x6)Au surface and the relative conductance (lower panel) during growth of
 Pb on Si(111)-(6x6)Au surface with annealed Pb layers. The letters A,B,C,D and E correspond to the coverage
 0.2, 1.2, 1.8, 3.2 and 4.2ML, respectively, at which deposition was terminated and Pb was annealed.}
\end{figure}
Fig. \ref{Fig1} shows the RHEED specular beam intensity changes during Pb deposition at 105 K on  Si(111)-(6x6)Au surface at a
glancing angle $0.5^o$ in the  azimuth of Si(111) surface (upper panel). The letters A,B,C,D and E mark the coverage at which
deposition of Pb was terminated. After annealing up to 300K and cooling down to 100K the conductance increases due to improved
crystalline order of the layer. For example in the case of the substrate Si(111)-(6x6)Au with deposited 4.2ML Pb after annealing
the conductance increases by $40\%$ at 100K. Moreover, the conductance of the substrate Si(111)-(6x6)Au with annealed Pb is
about two times larger than the conductance of Si(111)-(6x6)Au substrate. On such prepared surface further deposition of Pb was
carried out. The coverage dependence of the relative conductance $G/G_0$ during deposition of Pb on different substrates is
shown in Fig. \ref{Fig1}, the lower panel. Here $G_0$ means the conductance of the system with annealed Pb atoms, i.e.
Si(111)-(6x6)Au/Pb. For the substrate Si(111)-(6x6)Au/$\leq1.2$ML Pb, i.e. (6x6)Au with less than 1.2ML of annealed Pb, the
relative conductance increases with increasing Pb coverage (cf. lines A,B). It suggests that annealed Pb atoms form small,
few-atom clusters which are situated in (6x6)Au cells, \cite{jal4}. In  such a case small amount of deposited Pb atoms cannot
disturb the conductance as the electron mean free path is few times smaller than for the bulk material. On the other hand, the
decrease of the conductance starts from the beginning of the Pb deposition on the substrate with recrystallized Pb coverage
higher than 1.2 ML (cf. lines C,D,E). The conductance reaches a minimum value and afterwards increases. In this case annealed Pb
atoms form rather smooth surface, especially for integer monolayer, and the electron mean free path is close to the bulk one.
Deposition of Pb atoms on such prepared structure causes the roughness of the surface to increase - and the conductance to
decrease. But for the higher coverage these added atoms form 1ML high clusters and cause the conductivity increases.

\begin{figure}[h]
\begin{center}
 \resizebox{0.99\columnwidth}{!}{
  \includegraphics{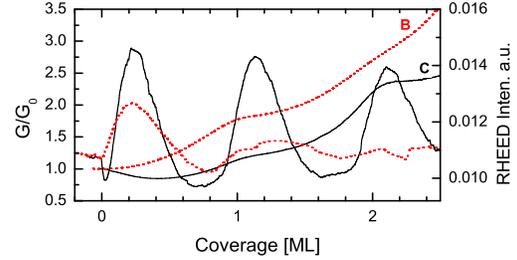}}
\end{center}
 \caption{\label{Fig2} (color online) RHEED specular intensity (right side) and relative conductance (left side)
 during growth of Pb on Si(111)-(6x6)Au surface
 with deposited and annealed  1.2 (broken lines) and 1.8ML (solid lines) of Pb, respectively.
 The relative conductance curves start from the value of 1 (for the coverage equals to 0). }
\end{figure}
The RHEED experiments confirm the above conclusions. In Fig. \ref{Fig2} we show the conductance (the lines start from the value
of 1 for the coverage equals to 0) and RHEED intensity of specular spot dependence on Pb coverage on Si(111)-(6x6)Au/1.2ML Pb
(broken lines) and Si(111)-(6x6)Au/1.8ML Pb (solid lines) (i.e. (6x6)Au with 1.2 and 1.8 ML of annealed Pb) surfaces,
respectively. In both cases the oscillations of the RHEED start from the beginning (layer-by-layer growth), but for the
substrate with recrystallized Pb coverage equal or lower than 1.2 ML these oscillations go out with the Pb coverage, cf. the
broken RHEED line. In this case the thickness dependence of the conductance is not monotonic. It should be noted that for the
substrate with more than 1.2ML of annealed Pb, the conductance oscillations (quantum size effect) have been observed.


\section{\label{sec4}Theoretical description}
To explain the behavior of the conductance of Si(111)-(6x6)/Pb during deposition of Pb atoms we have used the theory by Trivedi
and Aschroft \cite{tri}. The conductance of thin films in the presence of bulk impurity scattering and surface roughness
scattering can be expressed as follows:
\begin{eqnarray}
 \sigma_{xx}&=& {e^2k_F \over \hbar \pi^2} {1\over \kappa} \sum_{m=1}^{n_c} {{1-m^2/\kappa^2}\over
 {{2n_c+1 \over k_Fl_0\kappa}+\left({\delta d \over d}\right)^2{s(n_c)m^2 \over 3\kappa}}}
 \label{eq1}
\end{eqnarray}
where $k_F$ is the Fermi wave vector, $\kappa=k_Fd/\pi$, $l_0$ is the electron mean free path of the system, $n_c=Int(\kappa)$
and
\begin{eqnarray}
s(n_c)=(2n_c+1)(n_c+1)n_c/(3\kappa^3).
\end{eqnarray}
The function $\delta d$ is the root-mean-square deviation of the mean film thickness $d$. This function can be obtained from the
RHEED intensity oscillations during the growth of the layer \cite{jal1,coh15} and is given by \cite{horn}:
\begin{eqnarray}
(\delta d)^2=\sum_{n=0}^\infty (n-t/\tau)^2 (\Theta_n-\Theta_{n+1}) ,
\end{eqnarray}
 where $\tau$ is the deposition time of one layer (in calculations we set $\tau=1$) and $\Theta_n$ is the coverage of
the \emph{n}-th layer. Here we consider only the Pb coverage lower than 1ML (the second layer is forbidden) and in this case the
coverage function can be obtained analytically, $\Theta_{n+1}=1$ and $\Theta_{i>n+1}=0$, where $n$ is the number of full
substrate layers on which Pb atoms are deposited. Also the roughness function $(\delta d)^2$ has analytical parabolic form with
$\delta d(t=0)=\delta d(t=1)=0$ and the maximal value equals to 0.5 which corresponds to the half-coverage of the layer. In the
calculations we assume that the thickness of the considered system can be expressed in the following form:
\begin{eqnarray}
d=nd_0+td_0 ,\,\,\,\,\, t\in<0,1>,
\end{eqnarray}
where $t$ means the Pb coverage and $nd_0$ corresponds to annealed Pb layers (or Si(111)-(6x6)Au/Pb layers). We express all
distances in $d_0$ units and $k_F$ in $1/d_0$ units.
Moreover, we assume that for the substrate Si(111)-(6x6)Au/($<$ 1.2ML Pb) the electron mean free path is limited mainly by the
dimension of (6x6)Au cells and, as was deduced from our experiment, it is a few times smaller than the mean free path for the
substrate Si(111)-(6x6)Au/($>$1.2ML Pb).

\begin{figure}[h]
\begin{center}
 \resizebox{0.95\columnwidth}{!}{
  \includegraphics{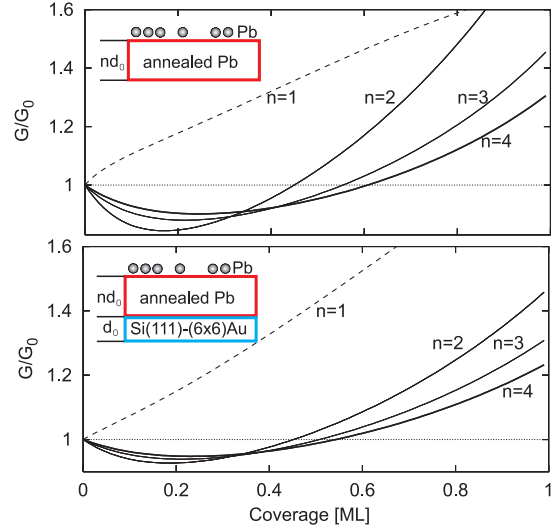}}
\end{center}
 \caption{\label{Figth} (color online) The relative conductance versus the coverage of Pb atoms (up to 1ML) on the surface consisted of $n-$annealed
 Pb layers (upper panel) and (6x6)Au with $n-$annealed Pb layers (lower panel). The mean free path is $l_0=25$ for $n=2,3,4$ and
 $l_0=7$ for $n=1$, $k_F=\pi$. To place the inside figure in the upper panel the broken line was divided by the factor $8$ and
 shifted to $1$.}
\end{figure}
In Fig. \ref{Figth} we show the relative conductance as a function of the Pb coverage for two kinds of substrates, i.e.
$n-$annealed Pb layers (upper panel) and (6x6)Au with $n-$annealed Pb layers (lower panel) - see inside figures. Here we
consider the same effective $d_0$ for (6x6)Au and for one Pb layer. It is interesting that for $n=1$ layer of annealed Pb the
relative conductance increases with increasing Pb coverage - for both substrates. In this case we set $l_0=7$, which e.g. for
$d_0=2.5$\AA  $ $ corresponds to the mean free path $l\simeq 18$\AA. For $n\geq 2$ and greater mean free path, $l_0=25$ (which
corresponds to the bulk one), the conductance decreases at the beginning and then for the higher coverage it increases. It is
worth mentioning that the minimum of the conductance appears when the electron mean free path is rather large (equal about the
mean free path of the bulk material). For very small $l_0$ this minimum does not appear.

The results obtained for the substrate Si(111)-(6x6)Au/Pb - the lower panel in Fig. \ref{Figth} - are in good agreement with the
experiment, cf. Fig. \ref{Fig1}, the lower panel. For the substrate with  $n$ annealed Pb layers - upper panel - this agreement
is somewhat worse as the curve for $n=1$ increases very rapidly (in Fig. \ref{Figth} this curve is divided by the factor 8) and
also the curve for $n=2$ increases too fast in comparison to experimental data.


\section{\label{sec5}Conclusions}
The electrical conductance at initial stage in epitaxial growth of Pb on the surface Si(111)-(6x6)Au/Pb (with up to 4.2 ML of
annealed Pb) has been measured using four-point probe. In comparison with similar experiments, e.g. \cite{pfe3}, where only the
conductance minimum was observed, our results show that for the substrate with lower than 1.2ML of annealed Pb  the conductance
increases with Pb coverage whereas for the substrate Si(111)-(6x6)Au with more than 1.2ML of annealed Pb the minimum in the
conductance appears. This effect can be understood in terms of electron mean free path of these substrates. For the substrate
Si(111)-(6x6)Au/$>$1.2ML Pb the mean free path is rather large, as in the bulk material and in this case the minimum in
conductance as a function of Pb coverage appears. In the case of the substrate Si(111)-(6x6)Au/$<$1.2ML Pb the electron mean
free path is a few times smaller than in the bulk Pb and the minimum in conductance does not appear - small amount of Pb atoms
cannot disturb the conductance. These adatoms cause the conductance to increase from the beginning of the epitaxy.  Moreover,
the RHEED intensity changes during the growth of the Pb layer show the difference for both surfaces. The oscillations of the
RHEED start from the beginning for both kinds of surfaces (layer-by-layer growth) but for the substrate with recrystallized Pb
coverage equal or lower than 1.2 ML these oscillations go out with the Pb coverage.


\noindent {\bf Acknowledgements} \\
The work of TK is partially supported by the Foundation for Polish Science. Authors thank Prof. M. Ja\l ochowski for helpful
discussions.


\end{document}